\documentclass{article}
\usepackage{fullpage}

\usepackage{amsmath}
\usepackage{amsthm} 
\usepackage{amssymb}

\usepackage{hyperref}
\usepackage{nonfloat}
\usepackage{rotating}
\usepackage{url,xspace}
\usepackage{bussproofs}
\usepackage{mathbbol}
\usepackage[all]{xy}
\SelectTips{cm}{}

\theoremstyle{plain}
\newtheorem{thm}{Theorem}[section]

\newcommand{\Const}{\mathit{I}} 
\newcommand{\Par}{\mathit{Par}} 
\newcommand{\Exc}{\mathit{Exc}} 
\newcommand{\inl}{\mathit{m}} 
\newcommand{\inr}{\mathit{n}} 
\newcommand{\try}[2]{\mathit{try}\{#1\}\,#2}  
\newcommand{\catchn}[1]{\mathit{catch}\,#1}  
\newcommand{\rais}[2]{\mathit{raise}_{#1,#2}}  
\newcommand{\handlen}[2]{#1\;\mathit{handle}\;#2}  

\newcommand{\Loc}{\mathit{Loc}} 
\newcommand{\Val}{\mathit{Val}} 
\newcommand{\St}{\mathit{St}} 
\newcommand{\prl}{\mathit{p}} 
\newcommand{\prr}{\mathit{q}} 

\newcommand{\id}{\mathit{id}}

\title{A duality between exceptions and states}
\date{October 24., 2011}
\author{
  Jean-Guillaume Dumas\thanks{
    LJK, Universit\'e de Grenoble, France. \url{Jean-Guillaume.Dumas@imag.fr}},
  Dominique Duval\thanks{
    LJK, Universit\'e de Grenoble, France. \url{Dominique.Duval@imag.fr}},
  Laurent Fousse\thanks{
    LJK, Universit\'e de Grenoble, France. \url{Laurent.Fousse@imag.fr}}, 
  Jean-Claude Reynaud\thanks{
    Malhivert, Claix, France. \url{Jean-Claude.Reynaud@imag.fr}}
}	

\begin{document}

\maketitle

\begin{abstract}
\textbf{Abstract.} 
In this short note we study the semantics of two
basic computational effects, exceptions and states,  
from a new point of view. 
In the handling of exceptions we dissociate the control 
from the elementary operation which recovers from the exception.
In this way it becomes apparent that there is a duality,
in the categorical sense, between exceptions and states. 
\end{abstract}

\section*{Introduction}
In this short note we study the semantics of two
basic computational effects, exceptions and states,  
from a new point of view. 
Exceptions are studied in Section~\ref{sec:exceptions}. 
The focus is placed on the exception ``flags''  
which are set when an exception is raised 
and which are cleared when an exception is handled. 
We define the \emph{exception constructor} operation
which sets the exception flag, 
and the \emph{exception recovery} operation
which clears this flag. 
States are considered in the short Section~\ref{sec:states}.
Then in Section~\ref{sec:dual} we show that 
our point of view yields a surprising result: there exists 
a symmetry between the computational effects of exceptions and states, 
based on the categorical duality between sums and products.
More precisely, 
the lookup and update operations for states are respectively 
dual to the constructor and recovery operations for exceptions. 
This duality is deeply hidden, 
since the constructor and recovery operations for exceptions 
are mixed with the control. 
This may explain that our result is,
as far as we know, completely new. 

States and exceptions are \emph{computational effects}:  
in an imperative language there is no type of states, 
and in a language with exceptions 
the type of exceptions which may be raised by a program 
is not seen as a return type for this program. 
In this note we focus on the denotational semantics 
of exceptions and states, so that the sets of states and exceptions 
are used explicitly. 
However, with additional logical tools, the duality 
may be expressed in a way which fits better with the syntax of effects 
\cite{DDFR10}. 

Other points of view about computational effects,
involving monads and Lawvere theories, 
can be found in \cite{Mo91,SM04,Le06,PP09}. 
However it seems difficult to derive from these approaches
the duality described in this note. 

\section{Exceptions}
\label{sec:exceptions}

The syntax for exceptions heavily depends on the language. 
For instance 
in ML-like languages there are several exception \emph{names}, 
and the keywords for raising and handling exceptions are 
\texttt{raise} and \texttt{handle},
while in Java there are several exception \emph{types}, 
and the keywords for raising and handling exceptions are 
\texttt{throw} and \texttt{try-catch}. 
In spite of the differences in the syntax, 
the semantics of exceptions share many similarities. 
A major point is that there are two kinds of values:
the ordinary (i.e., non-exceptional) values and the exceptions.
It follows that the operations may be classified 
according to the way they may, or may not, interchange 
these two kinds of values: 
an ordinary value may be ``tagged'' for constructing an exception, 
then the ``tag'' may be cleared in order to recover the value.

First let us focus on the raising of exceptions.  
Let $\Exc$ denote the set of \emph{exceptions}. 
The ``tagging'' process can be modelled by 
injective functions $t_i:\Par_i\to\Exc$
called the \emph{exception constructors},
with disjoint images: 
for each index $i$ in some set of indices $\Const$,
the exception constructor $t_i:\Par_i\to\Exc$
maps a non-exceptional value (or \emph{parameter}) $a\in\Par_i$ 
to an exception $t_i(a)\in\Exc$. 
When a function $f:X\to Y+\Exc$ \emph{raises} (or \emph{throws}) 
an exception of index $i$, the following \emph{raising} operation is called:
  $$ \rais{i}{Y}:\Par_i\to Y+\Exc $$ 
The raising operation $ \rais{i}{Y}$ is defined as 
the exception constructor $t_i$ followed by 
the inclusion of $\Exc$ in $Y+\Exc$.

Given a function $f:X\to Y+\Exc$ and an element $x\in X$, 
if $f(x)=\rais{i}{Y}(a)$ for some $a\in\Par_i$
then one says that $f(x)$ \emph{raises an exception of index $i$ 
with parameter $a$ into $Y$}.
One says that a function $f:X+\Exc \to Y+\Exc$ \emph{propagates exceptions} 
when it is the identity on $\Exc$. 
Clearly, any function $f:X\to Y+\Exc$ 
can be extended in a unique way as a function which propagates exceptions. 

Now let us study the handling of exceptions.
The process of clearing the ``exception tags'' can be modelled by  
functions $c_i:\Exc\to \Par_i+\Exc$ 
called the \emph{exception recovery} operations: 
for each $i\in\Const$ and $e\in\Exc$ 
the exception recovery operation $c_i(e)$ 
tests whether the given exception $e$ is in the image of $t_i$. 
If this is actually the case, 
then it returns the parameter $a\in\Par_i$ such that $e=t_i(a)$, 
otherwise it propagates the exception $e$. 

For handling exceptions of indices $i_1,\dots,i_n$ 
raised by some function $f:X\to Y+\Exc$, 
one provides a function $g_{i_k}:\Par_{i_k}\to Y+\Exc$,
which may itself raise exceptions, 
for each $k$ in $\{1,\dots,n\}$. 
Then the handling process builds a function 
which propagates exceptions,
it may be named $\try{f}{\catchn{i_k\{g_k\}_{1\leq k\leq n}}}$
or $\handlen{f}{(i_k\!\!\Rightarrow\!\! g_k)_{1\leq k\leq n}}$: 
  $$ \handlen{f}{(i_k\!\!\Rightarrow\!\! g_k)_{1\leq k\leq n}} : X+\Exc\to Y+\Exc $$
Using the recovery operations $c_{i_k}$, 
the handling process can be defined as follows.
\begin{tabbing}
 999 \= 999 \= 999 \= 999 \= \kill 
 \> For each $x\in X+\Exc$, 
   $\;(\handlen{f}{(i_k\!\!\Rightarrow\!\! g_k)_{1\leq k\leq n}})(x) \in Y+\Exc$ 
   is defined by: \\
 \>\> // \textit{if $x$ was an exception before the $try$,
     then it is just propagated} \\
 \>\> if $x\in \Exc$ then return $x\in \Exc \subseteq Y+\Exc$; \\
 \>\> // \textit{now $x$ is not an exception} \\ 
 \>\> compute $y:=f(x) \in Y+\Exc$; \\
 \>\> if $y\in Y$ then return $y\in Y \subseteq Y+\Exc$; \\
 \>\> // \textit{now $y$ is an exception} \\ 
 \>\> for $k=1..n$ repeat \\ 
 \>\>\> compute $y:=c_{i_k}(y)\in \Par_{i_k}+\Exc$;  \\
 \>\>\> if $y\in \Par_{i_k}$ then return $g_k(y)\in Y+\Exc$; \\
 \>\> // \textit{now $y$ is an exception but it does not have index  
  $i_k$, for any $k\in\{1,\dots,n\}$ } \\
 \>\> return $y\in \Exc \subseteq Y+\Exc$. 
 \end{tabbing}

Given an exception $e$ of the form $t_i(a)$, 
the recovery operation $c_i$ returns the non-exceptional value $a$
while the other recovery operations propagate the exception~$e$. 
This is expressed by the equations~(\ref{eq:exceptions-explicit})
in Figure~\ref{fig:exceptions}.
Whenever $\Exc=\sum_{i\in\Const}\Par_i$ 
with the $t_i$'s as coprojections, 
then equations~(\ref{eq:exceptions-explicit})
provide a characterization of the operations~$c_i$'s. 

\begin{figure}[!h]
\hrulefill
\begin{itemize}
\item[] For each index $i\in\Const$ :
\item a set $\Par_i$ (parameters)
\item two operations $t_i:\Par_i\to\Exc$ (exception constructor)
\\ and $c_i:\Exc\to\Par_i+\Exc$ (exception recovery) 
\item and two equations:
  \begin{equation} 
  \label{eq:exceptions-explicit} 
  \begin{cases} 
  \forall\,a\in \Par_i \,,\;
    c_i(t_i(a)) = a\in \Par_i \subseteq \Par_i+\Exc & \\
  \forall\,b\in \Par_j \,,\;
    c_i(t_j(b)) = t_j(b)\in \Exc \subseteq \Par_i+\Exc
    & \mbox{ for every } j\ne i \in \Const \\  
  \end{cases} 
  \end{equation} 
which correspond to commutative diagrams, 
where $\inl_i$ and $\inr_i$ are the injections: 
$$ \xymatrix@R=1pc@C=3pc{
\Par_i+\Exc & \Par_i \ar[l]_{\inl_i} \\
\Exc \ar[u]^{c_i} & \Par_i \ar[l]^{t_i} \ar[u]_{\id} \ar@{}[ul]|{=} \\
}  \qquad \qquad 
\xymatrix@R=1pc@C=2pc{
\Par_i+\Exc & \Exc \ar[l]_(.3){\inr_i}
  & \Par_j \ar[l]_{t_j} \\
\Exc \ar[u]^{c_i} && \Par_j \ar[ll]^{t_j} \ar[u]_{\id} \ar@{}[ull]|{=} \\
}  $$
\end{itemize} 
\hrulefill
\caption{Semantics of exceptions: 
constructor and recovery}
\label{fig:exceptions}
\end{figure}

\section{States}
\label{sec:states}

Now let us forget temporarily about the exceptions 
in order to focus on the semantics of an imperative language. 
Let $\St$ denote the set of \emph{states}
and $\Loc$ the set of \emph{locations}
(also called \emph{variables} or \emph{identifiers}). 
For each location $i$, 
let $\Val_i$ denote the set of possible \emph{values} for $i$.
For each $i\in\Loc$ there is a \emph{lookup} operation $l_i:\St\to\Val_i$ 
for reading the value of location $i$ in the given state.
In addition, for each $i\in\Loc$ there is
an \emph{update} operation $u_i:\Val_i\times\St\to\St$ for  
setting the value of location $i$ to the given value,
without modifying the values of the other locations in the given state. 
This is summarized in Figure~\ref{fig:states}. 
Whenever $\St=\prod_{i\in\Loc}\Val_i$ 
with the $l_i$'s as projections,
two states $s$ and $s'$ are equal 
if and only if $l_i(s)=l_i(s')$ for each $i$, 
and equations~(\ref{eq:states-explicit})
provide a characterization of the operations $u_i$'s. 

\begin{figure}[!h]
\hrulefill
\begin{itemize}
\item[] For each location $i\in\Loc$ :
\item a set $\Val_i$ (values) 
\item two operations $ l_i:\St\to\Val_i$ (lookup) 
\\ and $u_i:\Val_i\times\St\to\St$  (update) 
\item and two equations:
  \begin{equation} 
  \label{eq:states-explicit} 
  \begin{cases} 
  \forall\,a\in \Val_i \,,\; \forall\,s\in \St\,,\;  
     l_i(u_i(a,s)) = a & \\
  \forall\,a\in \Val_i \,,\; \forall\,s\in \St \,,\;   
     l_j(u_i(a,s)) = l_j(s) & \mbox{ for every } j\ne i \in \Loc \\  
  \end{cases} 
  \end{equation} 
which correspond to commutative diagrams, 
where $\prl_i$ and $\prr_i$ are the projections: 
$$ \xymatrix@R=1pc@C=3pc{
\Val_i\times\St \ar[r]^{\prl_i} \ar[d]_{u_i} & \Val_i \ar[d]^{\id} \\
\St \ar[r]_{l_i} & \Val_i \ar@{}[ul]|{=} \\
}  \qquad\qquad 
\xymatrix@R=1pc@C=2pc{
\Val_i\times\St \ar[r]^(.7){\prr_i} \ar[d]_{u_i} & 
  \St \ar[r]^{l_j} & \Val_j \ar[d]^{\id} \\
\St \ar[rr]_{l_j} && \Val_j \ar@{}[ull]|{=} \\
}  $$
\end{itemize} 
\hrulefill
\caption{Semantics of states: lookup and update}
\label{fig:states}
\end{figure}

\section{Duality}
\label{sec:dual}

Our main result is now clear 
from Figures~\ref{fig:exceptions} and~\ref{fig:states}.

\begin{thm}
\label{theo:duality}
The duality between categorical products and sums 
can be extended as a duality 
between the semantics of the lookup and update 
operations for states on one side 
and the semantics of the constructor and recovery 
operations for exceptions on the other side.
\end{thm}

In \cite{PP02} an equational presentation of states is given, 
with seven families of equations. 
These equations can be translated in our framework,
and it can be \emph{proved} that they are equivalent
to equations~(\ref{eq:states-explicit}) \cite{DDFR10}.
Then by duality we get for free seven families of equations 
for exceptions.
For instance, it can be proved that 
for looking up the value of a location~$i$ 
only the \emph{previous} updating of this location~$i$ is necessary, 
and dually, 
when throwing an exception constructed with $t_i$
only the \emph{next} recovery operation $c_i$, with the same index~$i$, 
is necessary. 

\bibliographystyle{alpha}

\begin{thebibliography}{}

\end{thebibliography}


\begin{thebibliography}{99}

\bibitem[Dumas et al. 2010]{DDFR10} 
Jean-Guillaume Dumas, Dominique Duval, Laurent Fousse, Jean-Claude Reynaud.
States and exceptions considered as dual effects. arXiv:1001.1662 v4 (2010). 

\bibitem[Levy 2006]{Le06}
Paul Blain Levy.
Monads and adjunctions for global exceptions. 
MFPS 2006. 
Electronic Notes in Theoretical Computer Science 158, p.~261-287 (2006).

\bibitem[Moggi 1991]{Mo91}  
Eugenio Moggi.
Notions of Computation and Monads.
Information and Computation 93(1), p.~55-92 (1991).

\bibitem[Plotkin \& Power 2002]{PP02} 
Gordon D. Plotkin, John Power.
Notions of Computation Determine Monads. 
FoSSaCS 2002.
Springer-Verlag 
Lecture Notes in Computer Science 2303, p.~342-356 (2002).

\bibitem[Plotkin \& Pretnar 2009]{PP09} 
Gordon D. Plotkin, Matija Pretnar.
Handlers of Algebraic Effects. 
ESOP 2009.
Springer-Verlag 
Lecture Notes in Computer Science 5502, p.~80-94 (2009).

\bibitem[Schr{\"o}der \& Mossakowski 2004]{SM04} 
Lutz Schr{\"o}der, Till Mossakowski.
Generic Exception Handling and the Java Monad. 
AMAST 2004.
Springer-Verlag 
Lecture Notes in Computer Science 3116, p.~443-459 (2004).

\end{thebibliography}

\end{document}